# Physical Foundations of Consciousness
# Brain Organisation:  The Role of Synapses

Charles T. Ross and Shirley F. Redpath.

**Abstract**

*We have analysed the many facets of Consciousness into two distinct categories. First: the organisational state of the neural networks at any one time, which determines whether a person is conscious – awake, or unconscious – asleep. Second: the processes that underlie the traffic of electrical signals across these networks that accounts for all the experiences of conscious awareness. This paper addresses the former; namely, how the state of the billions of neural networks and the trillions of additional axons, dendrites and synapses varies over the daily cycle - what physically changes when we go to sleep – what happens when we wake up.*

*We submit that the widths of synaptic clefts are not fixed, but are variable, and that this variable tension across the synapses is the neural correlate of consciousness.*

**State and Process**

The many attributes of Consciousness can be usefully divided into two distinct categories: the *state* of the mass of the neural networks; and the *processes* that underlie the traffic transmitted over those networks: the organisational state of the brain as opposed to its cognitive functions.

The former determines the current experience of conscious awareness along the gradient from deep anaesthesia through simple unconscious sleep to being consciously awake, alert and all the way to full concentration.  The latter includes all the many familiar activities of monitoring and processing the sensations of the world. It includes all the processing associated with being able to see, hear, feel, taste and touch, and to move speak, learn, think and create new ideas and concepts. It also includes all the neural activity generated by the emotions, feelings, attitudes, opinions and the sense of having a degree of control over behaviour. It is possible to observe from brain scans that activity takes place over an individual's neural networks when they are in a state of being both conscious and unconscious; however, in the latter state they are not, by definition, consciously aware of that activity, and, indeed, they only ever will be aware of that activity when, and if they wake up.

This problem of consciousness has been facing - and defeating - scientists for many generations. What happens in our brains when we wake up? It seems such a simple question, but no one has ever come up with a plausible and coherent answer. In a paradigm where everyone is stuck, a fresh set of ideas is always a step forward, and we believe that by separating the state of the neural networks from the processes of the traffic over these networks, and then concentrating on understanding the former, it will be much easier to begin to understand how all the aspects of consciousness fit into the jig saw.

This paper, therefore, addresses the former; namely, how the state of the billions of neural networks and the trillions of additional axons, dendrites and synapses varies over the daily cycle - what changes when we go to sleep – what happens when we wake up.

The experience of 'being conscious' – awake – is very different from 'being unconscious' – asleep. At one extreme is to be alert and concentrating hard. At the opposite extreme is anaesthesia. Deep sleep equates with being unconscious. Being wide awake equates with being conscious, but in addition there is an underlying subconscious state that is continuously in operation. The senses continue to monitor the outside world, even if there is no conscious experience of this activity. If someone's name is called out, even very quietly into their sleeping ear, they awake. A shouted warning or the sounding of an alarm will wake people. It is not the volume, but the content of the message that is important. The brain must be receiving and processing everything that is going on around it to achieve this feat.

Similarly, although the subconscious monitors all the sensory activity in its environment, this information may not penetrate into conscious awareness even when the person is fully awake, because the relative strength of the signal is not sufficiently strong to need a response.

**Three States**

Thus, we can observe three states. *Consciousness* – being awake. *Unconsciousness* – being asleep: and, permanently in the background, the *Subconscious*. Over the centuries, many observers have tried to explain this third state, but it was Freud[1] and Jung[2] around a century ago, who began to identify and delineate the subconscious. We can note that some Hindus and Buddhists would argue that there is a fourth, or 'super conscious' state.

It is possible to be asleep yet be dreaming so vividly that the sense of involvement is indistinguishable from being consciously awake – during the dream, at least. It is also possible to be fully awake yet daydreaming happily so as to be in a state similar to sleep. Consciousness is often associated with being aware of all the sensations of the surrounding world, yet it is possible to wake up in, say, a dark room, with no sounds, no smells, no tastes and only the slightest touch of some bedclothes – e.g. to have no physical stimuli – yet be fully conscious, able to imagine complex scenarios of potential events and feeling very much alive and in control.

We appear not to have conscious control over our transition from one state of consciousness to another. It is possible to lie in bed unable to go to sleep for hours on end, but a sharp blow to the head can make someone instantly unconscious. It is possible to fight off drowsiness or sleep for limited periods to cope with some emergency situation, but eventually sleep overtakes everyone. Sustained emotional activity, which uses up a lot of energy, invariably causes sleep.

**Separation of State From Process**

All these examples, and many others, reinforce the argument that being conscious or unconscious is separate from whatever mental activities are, or are not, going on at any one moment. In short, whenever someone is 'conscious' they experience a steady flow of thoughts, ideas, concerns, intentions, reactions and feelings as they monitor and relate to the environment or give free rein to their imaginations. Equally, people can discipline themselves to slow down all that mental activity, shut out the outside world and meditate, which is possibly the basis of a super conscious state. It is very easy to slip from imagination, a conscious process, to dreaming, an unconscious one. If we can isolate all the processes of consciousness that generate the mass of electrical signals over the neural networks of the brain we are left with the 'state' of those networks.

More has been discovered about the neural networks in the last few decades than in all previous recorded history. A remarkable amount is known about the neurons, their nuclei and the axons and dendrite filaments that connect them to the senses, the muscles, the glands and all the other organs of the body, thereby linking and coordinating every organ in the body with every other. We know that there are more glia cells than neurons, although we are not fully aware of their myriad functions. Among a number of power generating systems, mitochondria in the nuclei create energy from the nutrients circulating in the blood stream. Action potentials are generated along the axons and dendrites. The whole brain is bathed in a myriad of chemicals; some that bind to the axons and dendrites to modify specific networks, and others that broadcast general messages.

**Synapses**

One aspect that has attracted particular attention is the way in which all neurons are connected to each other. Neurons do not physically touch; instead, they communicate across tightly controlled gaps between their respective axons and dendrites. These junctions, known as *synapses* from the Greek word 'to clasp', are surprisingly sophisticated. There is a narrow gap, or cleft, between the transmitting and receiving terminals. In principle, when an action potential message flowing along an axon reaches a synapse, it causes various neurotransmitters to be excreted. These swim across the synaptic cleft and bind to the receiving dendrite, initiating a corresponding action potential which enables the message to continue its journey. Synapses can pass messages, amplify messages and negate messages. The pharmaceutical industry has been very active in research on the complex operation of synapses and many preparations like valium™ and prozac™ operate on the chemicals that are believed to operate in and around the synapses. These drugs may variously stimulate the production of neurotransmitters, inhibit them, operate on hormones that affect transmissions along the axons and dendrites, or seek to affect excess neurotransmitters that in normal circumstances would be retrieved by the axon terminals that over supplied them. Chemical warfare agents, such as sarin™, are particularly lethal because they interfere with synaptic links.

There is another aspect of synapses that is less well understood. Not much is known about the width of synaptic clefts. It is not known what holds the two sides together, in close proximity but without touching. There is, as yet, no plausible theory about how or why this method of connecting neurons together has evolved in such a curious and immensely complex way, yet we know that Nature rarely evolves without a very good reason.

**Hypothesis**

*We propose the hypothesis that the widths of synaptic clefts are not fixed, but are variable, and that this variable tension across the synapses is the root operand of the various states of consciousness.*

**How might this operate?**

Observing someone gradually falling asleep it is noticeable that their muscles relax, their eyes close and breathing drops to a steady rhythm. We propose that similar variations in the amount of energy directed to the synapses causes variations in the tension across the synaptic clefts. As these energy levels drop, the synaptic gaps gradually widen. This does not completely inhibit the transmission of messages, but will slow them down and make that transmission less efficient. Below a critical point, the volume of perceived activity falls below a horizon and the person drifts into a state of unconscious sleep. After a period of rest, the energy levels in the neurons are gradually replenished, increasing the tension across the synapses, closing the synaptic gaps and facilitating the efficient passage of neural activity. When a critical mass is reached, the volume of perceived activity will rise above the horizon and the person becomes consciously awake.

*Thus the level of tension across the neural networks determines the* state of consciousness. *The type and volume of traffic over those networks determines the* experiences of consciousness.

**How might this have evolved?**

As groups of primeval cells banded together to create larger organisms, the earliest brains evolved to link together all the developing senses, muscles and organs, enabling an organism to manage its behaviour and operate as one co-ordinated, co-operative whole in a process known as the autonomic system. This neural activity uses up a lot of energy. However, being able to respond increasingly quickly to threatening stimuli, proved to be an extremely valuable attribute in the battle for survival. Energy is always a scarce resource, and the conservation of energy has always been a major constraint on living organisms. Life forms that could deploy increasing amounts of neural energy, even if only over short periods, prospered. Thus, the underlying autonomic systems evolved two new operating states. Neuron networks began to deploy unsustainable amounts of energy over short periods, followed by periods of relaxation to enable this energy to be replenished. How could the same neural networks operate in these two different modes?

We propose that synapses initially evolved as simple 'circuit breakers'. However, the ability to stay connected but vary the tension holding synapses together and so vary the width of the gaps across the synaptic clefts provided as increasingly sophisticated, efficient and flexible system. In the active phase tension was increased and the synaptic cleft was minimised, enabling the neuron networks to operate at maximum efficiency even at the cost of incurring unsustainable energy usage. As the available energy dropped the tension across the synapses reduced, widening the synaptic clefts and causing the traffic across the network to fall. This reduced the energy usage to the point where the neurons could generate more energy than was being used and so replenish their stocks. The first condition developed into the state we recognise as conscious awareness – being awake and able to respond to any eventuality. The latter developed into the state of relaxation and inactivity we recognise as being unconscious – of being asleep. Because the synaptic cleft was not completely severed, the autonomic system was able to continue operating in the background ensuring the survival of the organism.

This tripartite system – the *subconscious* state managing the underlying, autonomic or basic 'operating system' – the fast reacting self controlled *conscious* state – and the compensating relaxation and recharging *unconscious* state, proved to be so successful that it has been one of the principal drivers of evolution. As the conscious state has evolved it has demanded ever more energy, reinforcing the mechanism.

The energy generation mechanism in cold-blooded reptiles will only allow them to sustain full conscious awareness in direct sunlight. Animals that pushed the envelope of conscious awareness outside the margins of daylight prospered. They developed warm blooded strategies to store energy for use in darkness. This used more energy, in turn requiring more effective food gathering strategies such as working in groups, developing tools and communicating through language. The human brain, employing the most complex mix of these strategies, uses one third of all the energy generated by the whole body.

**Imagination and Dreaming**

As people drift off to sleep their motor neuron synapses tend to lose tension and so drift apart first. Humans curl up and lie down. Next to lose tension tend to be the networks associated with sensory awareness of the surroundings. This leaves the neuron to neuron links, which control our language, thinking and imagination systems. These links use the least amount of energy, so they remain active the longest. In this state it is possible to continue imagining, but when the control systems lose tension and relax this imagination slips into dreaming. Finally those networks lose tension and the whole brain is asleep, with the exception of the autonomic system.

Several observed phenomena of the sleeping state demonstrate the delicate balance between state and process. The decoupling of the motor neurons explains the frequently reported dream sequence where people are aware of some danger but have the sensation that they cannot move out of the way. Conversely, in a minority of circumstances the motor neurons may remain active after other networks relax, which accounts for the phenomena of sleep walking. Similarly, people talk in their sleep when only the language networks have remained active. People report being conscious of, say, a radio transmitting a discussion while they are otherwise asleep and feeling frustrated that the participants do not let them join in. It appears that the body is asleep, yet the brain is continuing to run some processes.

**Prioritising**

This hypothesis of synaptic tension directly enabling the performance of neural networks also accounts for another attribute of consciousness. It is possible to carry out more than one task at a time and to switch attention very quickly from one subject to another. For instance, people drive cars while concentrating on a conversation with their passengers, but if a dangerous event happens on the road in front of them, the driver instantly directs full concentration to the task of driving. In escalating crises the brain can switch its resources to what appears to be the most immediate priority very swiftly by pumping energy into the target neural network. This raises the tension across the synapses of that network, closing the clefts and allowing messages in that area to transmit faster and stronger than in other areas of the brain. By this means one whole set of networks can take over from another very quickly.

*Thus it can be seen that synaptic tension not only determines the conscious state of the brain but also the level of conscious awareness and activity, thereby setting the priority of all its response systems.*

There is an example of this from the burgeoning technology of robotics. There is a design issue in robotics referred to as the 'concurrency problem'. Put in simple terms, a robot may walk over a cliff, not because its sensors have not noticed the void in front, but because it's processing resources are devoted to another task and can not respond fast enough. In the biological world, survival depended upon the ability of one group of neurons to be able to grab the attention of the whole system instantly, so that they could concentrate on coping with just such an emergency.

**Theory Applications**

Concussion

A blow to the head actually knocks the synapses apart causing instant unconsciousness. As the brain works to regain the normal level of tension in the neural networks the clefts close up and consciousness is gradually regained. Frequent concussion will weaken this ability to repair the system.

Variable Memory Loss

Variable synaptic gaps may account for another well recognised phenomenon of the brain. When trying to recall some piece of information people often report a sense that they know the answer but that it is just out of reach – it is on the 'tip of their tongue'. The reason may be that the networks representing that information are activated, but the strength of the signal is not strong enough to cross intervening synaptic clefts, and therefore it is just out of reach. People similarly report that in a crisis they can recall all sorts of 'lost' information, like important telephone numbers. The energy generated by such a crisis closes the synaptic clefts enabling weak signals to pass and thus the information may be recalled.

It is probable that some aspects of dementia are caused by a weakening of the system to generate the levels of tension necessary to facilitate the passage of neural signals across synaptic clefts. As they drift apart, less and less neural activity penetrates into consciousness.

Death and Brain Damage

When the body dies, all electrical activity ceases. All the muscles lose their tension and so the organs cease to function. All neural activity ceases. Temporary links will disintegrate. All the synapses fall apart and the subconscious as well as consciousness ceases to exist. For these reasons it is very difficult to observe the workings of a human brain, because it significantly changes its configuration at death. It is also difficult to observe the operations of a live brain in this detail, but we can infer certain neural activities from our observations of localised brain damage. A stroke causes a partial failure of sections of the brain. Depending on the severity of the stroke, some networks lose their tension, and therefore the ability to recall memories, either permanently or for a period of time. Through its incredible resilience, the brain often gradually restores tension in the affected networks, facilitating the transmission of traffic, and so restoring memory.

Synchronisation

Recent research using brain scanners has shown that large groups of networks of neurons can synchronise their electrical activity and fire at the same frequency; a process known as 'phase locking'. By using different frequencies, these groups of networks are able to operate concurrently without interfering with one another[3]. Whichever group of networks is receiving the most energy is in the highest tension and so seizes conscious attention. There is some evidence of energy sweeping in waves across brain areas. This might be an example of how the brain ratchets up synaptic tension.

**Implications**

We now have a plausible hypothesis to explain the physical manifestations that occur as we move through all the different states of our conscious awareness - from deep anaesthesia, though sleep and waking up, to full concentration - and how our attention is directed to what appears to be the highest priority.

What this hypothesis is attempting to answer is what the Australian philosopher David Chalmers calls the 'hard problem': determining how physiological events in the brain translate into what we experience as consciousness, or, to put it another way, how a neural phenomenon causes a human experience[4].

*Synaptic tension is the neural correlate of consciousness.*

Baroness Susan Greenfield, Professor of Pharmacology at Oxford University, argues that, while the size of assemblies of neurons that are active at any one time do not create consciousness, the size of these assemblies are indices of degrees of consciousness[5]. On the other hand, Christof Koch, who worked with Francis Crick, joint discoverer of the structure of DNA, and is now Professor of Cognitive and Behavioural Biology at the California Institute of Technology, suggests that it is the informational complexity of arrays of active neurons that is significant, i.e. it is quality, not quantity that determines consciousness[6].

*There is a difference between being conscious, and being conscious* **of** *something*.

Thus, the size of neuron assemblies and their informational complexity affect what a person is conscious of at any one time. If, however, the networks are in resting state – unconscious – neither the size nor complexity of the neuron activity will have any impact.

What matters is the amount of energy being deployed. But again, this is at two levels. The energy pumped into the neurons to raise the tension across the synapses to increase the efficiency of the whole network determines consciousness; and the strength of the signals transmitted across these networks determines what the brain is conscious of. Both the Greenfield and Koch conjectures apply to the latter. There is a lot to be learned from this line of discussion.

Habituation

Many activities impinge on the conscious self only if something out of the ordinary happens to grab the attention. Each time a non-significant or routine event is repeated the neural energy needed to process its response instructions is reduced, so eventually it does not reach the critical mass needed to penetrate consciousness. We use the term 'habituation' for this effect, which can be observed many times each day. People report that frequently performed activities like walking, driving, washing the dishes, or getting dressed in the morning appear to be performed as though on 'automatic pilot'.

Self Control

It is possible to deliberately direct attention. In most circumstances it is possible to choose which sensory input to concentrate upon. It is possible to decide what to think about, to examine desires, thoughts and ambitions, and imagine other situations. It is possible to direct concentration selectively. It is possible to listen to just one instrument in an orchestra or concentrate on the conversation of one person in a crowded, noisy room. It is possible to zoom in on one visual image, taste, smell or feeling. Coherent streams of words can be pre-assembled to conduct conversations with others. Similarly, people can conduct a reasoned stream of thoughts with themselves. Conceptually, it is possible to hold a debate across the corpus collosum. These are all examples of neural activity over the networks when they are in a high tension, conscious state.

At the basic level, underpinning the whole of existence, the autonomic brain is contributing substantially to the co-ordination and smooth running of the whole body. What goes on is registered, but relatively little impinges on awareness. Only very rarely are these activities 'conscious' and that is generally if something goes wrong. Hardly ever is anyone aware of the body's ongoing maintenance work to remove, mend or replace damaged or worn out tissues. The immune system is permanently active yet most people are largely oblivious of it.

**Summary**

Conscious awareness requires two conditions. The networks must be in a conscious state and the activity over those networks must also be sufficiently strong to attract attention. If the networks are in an unconscious state no amount of activity over the networks will attract conscious attention, but in certain circumstances subconscious activity may be sufficient to cause energy to be pumped into the networks to raise them to a conscious state; a shouted warning, for instance. If the networks are in a conscious state, considerable volumes of neural traffic may or may not attract conscious attention, depending on whether some other neural activity has grabbed attention and so is taking precedence.

Whether the neural networks are in a conscious or unconscious state is determined by the strength of the tension across the synapses and therefore the widths of the synaptic clefts.

Humans have evolved some degree of control over the activities of their brains, but this is observably incomplete. In a crisis, reaction becomes automatic: if physically confronted, either the fight or flee instinct takes over. In a heated argument responses may not be fully under conscious control. People report the experience of hearing themselves come out with some riposte that in less stressful circumstances they might not have used. Under pressure, there is only partial conscious control. There is no time to think of a more reasoned response. The brain deploys the maximum energy available to execute the swiftest response by the most direct means, by-passing all its more sophisticated facilities.

However, to varying degrees and depending on the circumstances, the ability has evolved to interrupt the instinctive or conditioned responses and delay taking action until alternatives have been considered. It is possible to select and initiate what to think about, how to think about it and what action to take. Imagination is, possibly the most valuable attribute – it is the basis of thinking and creativity. It is only possible for people to be imaginative when they are conscious, when they can initiate actions and have control over their imaginings.

This all strongly suggests that there are a series of layers of neural networks. There are neural networks that operate the automatic functions of the body, and behaviours such as the fleeing instinct. These are overlaid by more complex neural networks that monitor current and stored experience, and can interrupt the automatic response. Perhaps overlaying both of these, there are neural networks that monitor other neural networks and, through the process of imagination, create a whole new range of options. Through all these, a system has evolved that enables the identification and evaluation of alternatives, over which increasing levels of control can be exercised. Consciousness is experienced if the level of tension holding the synapses together is strong enough to enable the electrical activity across the networks to grab attention.

**Corroboration**

The next step is to measure the widths of synaptic clefts and then see by how much these values vary between the conscious and unconscious states. This is right on the edge of current capabilities. Electron microscopy can measure the width of axons and work has started on reconstructing circuit diagrams of sections of the brain, but this involves terabytes of data and advances in image analysis. Multiphoton microscopy, which uses fluorescence and ultrafast lasers, might allow us to identify regular variations in the synaptic clefts. Sam Wang, at Princeton, reports that his team has been able to measure the strength of signals across synapses, but not the ability to measure the width of the clefts as yet[7].

Perhaps confocal microscopy scanning being developed by Qinetiq PLC in the UK could be adapted to this task.

Based on work by Leon Chua at the University of California, Berkeley, Stan Williams' team at Hewlett Packard have devised an electronic component like a resistor that has a form of memory, which they have called a 'memristor'[8]. Max Di Ventra, a physicist at the University of California at San Diago, and Yuriy Pershin have built a transistor-memristor chip that appears to emulate synapses[9]. This work suggests that synapses may have evolved their physical attributes to be able to alter their response according to the frequency and strength of signal traffic. Thus synapses may contribute both to the process of the storage of information – memory - as well as determining the state of consciousness of the neural networks.

**Conclusion**

This model provides a plausible explanation of the varying states of the neural networks. It explains what actually happens when a person wakes up and becomes conscious of being alive and aware of everything going on about them; what makes them capable of imagining different situations and having some control over their own behaviour; and what happens as they drift off to sleep.

Proof of this model of the neural correlates of consciousness, will enable us to understand the manifold other aspects of consciousness that occur as a result of the mass of electrical activity that flows over these networks.

Charles T. Ross (Hon)FBCS, FIAP, CITP, FIMIS.
Charles.Ross@BrainMindForum.co.uk

Shirley F. Redpath MA, MBA.
Shirley.Redpath@BrainMindForum.co.uk

These hypotheses, ideas, theories, conjectures and their implications are set out in greater detail in *Biological Systems of the Brain: Unlocking the Secrets of Consciousness*, by Charles Ross & Shirley Redpath.
http://www.BrainMindForum.co.uk
ISBN  978 1848760 004  http://www.troubador.co.uk/book_info.asp?bookid=671

See also http://arxiv.org/abs/0905.2836 by the same authors.

See also http://www.cycognition.com